# Bell's Theorem refuted with a Kolmogorovian counterexample

J.F. Geurdes
C.vd. Lijnstraat 164
2593 NN Den Haag
Netherlands
geurd030@planet.nl

Abstract

The statistics behind Bell's inequality is demonstrated to allow a Kolmogorovian model of probabilities that recovers the quantum covariance. It shown that P(A=+1)=P(A=-1)=1/2 given A=1, or A=-1 is the result of measurement with device A. The occurrence of irregular integration prevents the use of Schwarz's inequality, hence, prevents the derivation of Bell's inequality. The obtained result implies that not all local hidden variables theories are ruled out by Aspect's (see Ref 4) experiment.

## 1. Introduction

In the early beginnings of quantum theory (QM) questions of interpretation arose. In a later stage, an important step in the debate was made by Bell[1].

Based on Einstein's criticism (EPR paradox) of completeness[2], Bell formulated an expression for the relation between distant (spin) measurements such as described by Bohm[3]. In Bell's expression, local hidden variables (LHV's) to restore locality and causality to the theory are introduced through a probability density function and through their influence upon the elementary measurement functions in the two separate wings (denoted by the A- and the B-wing) of the experiment.

Many experiments and theoretical developments arose from Bell's original paper. The most important experiment was performed by Aspect[4]. Aspect's results were interpreted as a confirmation of the completeness of quantum mechanics and the impossibility of local hidden causality. From that point onwards, QM was considered a non-local theory. In a previous paper, the present author has discussed this proof[5] arguing that there was insufficient ground for such a far reaching conclusion.

Let us describe briefly a typical Bell experiment. In this kind of experiment, from a single source, two particles with, e.g. opposite spin, are send into opposite directions. We could think of a positron and an electron arising from para-Positronium, drawn apart by dipole radiation. Subsequently, in the respective wings of the experiment, the spin of the individual particle is measured with a Stern-Gerlach magnet. The measurements are found to covary and depend on the, unitary, parameter vectors, $(a_1, a_2, a_3)$ and $(b_1, b_2, b_3)$ of the magnets. Quantum mechanics predicts the correlation to be

$$P(\vec{a}, \vec{b}) = -\sum_{k=1}^{3} a_k b_k \qquad (1)$$

Bell's theorem states that local hidden variables cannot recover P(a,b), because all local hidden distributions run into inequalities similar to the one stated below. Vectors a, b, c, d are unitary.

$$\left| P_{LHV}(\vec{a}, \vec{b}) - P_{LHV}(\vec{a}, \vec{d}) \right| + \left| P_{LHV}(\vec{c}, \vec{b}) + P_{LHV}(\vec{c}, \vec{d}) \right| \leq 2 \qquad (2)$$

In the present paper, Eq. (1) will be derived from local hidden variables. Obviously this cannot be done with regular integration.

## 2. Preliminary remarks

In this section the proposed model for a LHV explanation of the EPR paradox is introduced. The basis of the model is the Gaussian probability density.

When a Gaussian density is normal with a mean zero and a standard deviation of unity we write, N(0,1). The Gaussian density with mean zero, standard deviation, |T|, is

$$\delta_T(x) = \frac{1}{|T|\sqrt{2\pi}} \exp\left[-\frac{x^2}{2|T|^2}\right]$$  **(3)**

Hence, $\delta_T(x)=N(0,|T|)(x)$. This is equal to the Dirac delta, when $T\rightarrow 0$, T and x independent.

Associated to the previous function is the 'Heaviside' function which is

$$H_T(x) = \int_{-\infty}^{x} \delta_T(y)\, dy$$  **(4)**

From the fact that, $H_s(0)=1/2$, for all standard deviations, s>0, it follows that in the limit, $T\rightarrow 0$,

$$\lim_{T\rightarrow 0} H_T(x) = \begin{cases} 1, & x>0 \\ 1/2, & x=0 \\ 0, & x<0 \end{cases}$$  **(5)**

In the following the behaviour of the Heaviside function in zero needs to be altered. We therefore define an alternative form of Heaviside function.

$$\theta_T(x) = H_T(x) + 2H_T(x)H_T(-x).$$  **(6)**

In the limit, $T\rightarrow 0$, we have

$$\lim_{T\rightarrow 0} \theta_T(x) = \theta(x) = 1, \quad x\geq 0,$$
$$\lim_{T\rightarrow 0} \theta_T(x) = \theta(x) = 0, \quad x<0.$$  **(7)**

Furthermore, differentiation to, x, and taking the limit, $T\rightarrow 0$, gives the result

$$\lim_{T\rightarrow 0} \frac{d}{dx}\theta_T(x) = \delta(x) - 2\delta(x)\, sign(x).$$  **(8)**

Because, $sign(x)=H_T(x)-H_T(-x)$, for, $T\rightarrow 0$, and, $\delta_T(x)=\delta_T(-x)$, we see that

$$\delta_{alt}(x) = \lim_{T\rightarrow 0} \frac{d}{dx}\theta_T(x)$$  **(9)**

is unequal to zero only when x=0, and coincides with, $\delta(x)$, because, by definition, sign(x), is zero in x=0. Hence, the alternative Heaviside function in Eq. (6), has the proper behaviour. In view of the definition, $\delta_T(x) = N(0,|T|)(x)$, it is remarked that, $|T|\delta_T(x)$, lies in the interval [0,1]. Subsequently let us define the sum terms

$$\sigma_A = \sum_{k=1}^{3} a_k \, sign(y_k)$$

$$\sigma_B = \sum_{k=1}^{3} b_k \, sign(y_k)$$

$$|\sigma_A| \leq \sqrt{3}, \ |\sigma_B| \leq \sqrt{3}$$

**(10)**

with $a_1^2 + a_2^2 + a_3^2 = 1$, and, $b_1^2 + b_2^2 + b_3^2 = 1$, and the variables $y_k$, are normally distributed, $N(0,1)(y_k)$. The entities, $a_k$, and, $b_k$, represent the unitary directional parameters of the measuring devices.

3. Probabilistic behaviour of the measurement devices

In the model, a major role is played by two sets of each three variables carrying the index A or B. The indices refer to the measuring device in the left-hand (A) or the right-hand (B) wing. Here in the definition section the use of the indices can be suppressed because we assume that identical measuring devices interact identically with the to-be-measured entity.

In the first place, let us introduce the probability density for the T variable employed earlier. We have

$$\rho_T(T) = \begin{cases} \delta(T-0^+), & T \in [-\epsilon, \epsilon] \\ 0, & T \notin [-\epsilon, \epsilon] \end{cases}$$

**(11)**

with, $\epsilon$, a small positive real number. Here, the $0^+$ means the limit from above to zero which may come arbitrary close to zero, but staying unequal to zero at the same time. Hence, for an arbitrary otherwise well-behaved function, f, we have

$$\int_{-\infty}^{\infty} \rho_T(T) f(T) \, dT = \lim_{0 < T \to 0} f(T) = f(0)$$

**(12)**

Furthermore, let us define a distribution for a variable, $\beta$, such that,

$$\rho_B(\beta) = \delta(\beta - |T|), \qquad \beta \in \mathbb{R}.$$

**(13)**

Before we introduce the probability density of the x variable used previously, we note beforehand that

$$\lim_{0 < T \to 0} T^T = \lim_{0 < T \to 0} \exp[T \ln T] = \exp[\lim_{0 < T \to 0} T \ln T] =$$

$$\exp[\lim_{0 < T \to 0} \frac{1/T}{-1/T^2}] = \exp(0) = 1$$

**(14)**

This limit is crucial in the construction and evaluation of the model. The probability density (pdf) for x is then defined by

$$\rho_X(x; T, \beta) = \begin{cases} 1/|T|^{\theta(x)}, & x \in [-\frac{1}{2}|T|^\beta, \frac{1}{2}|T|] \\ 0, & x \notin [-\frac{1}{2}|T|^\beta, \frac{1}{2}|T|] \end{cases} \tag{15}$$

Note that the Heaviside (theta) function, $\theta(x)$, in the previous definition is equal to the one defined above, that is, $\theta(x)=1$, when $x \geq 0$, while, $\theta(x)=0$, when $x<0$. Note also that

Given all the previous definitions of this section, we can now come to the evaluation of the associated integral over, T, over, $\beta$, and, over, x. This integral will return in the model in relation to the A and B indexed variables. Suppressing those indices for this moment we have, for some auxiliary variable, $W_0$,

$$W_0 = \int_{-\infty}^{\infty} dT \int_{-\infty}^{\infty} d\beta \int_{-\infty}^{\infty} dx\, \rho_T(T)\, \rho_B(\beta)\, \rho_X(x; T, \beta) \ . \tag{16}$$

Evaluation of the integral over x and its density gives

$$\int_{-\infty}^{\infty} dx\, \rho_X(x; T, \beta) = \int_{-\frac{1}{2}|T|^\beta}^{\frac{1}{2}|T|} dx\, \frac{1}{|T|^{\theta(x)}} \tag{17}$$

This leads to the integral

$$\int_{-\infty}^{\infty} dx\, \rho_X(x; T, \beta) = \int_{-\frac{1}{2}|T|^\beta}^{0} dx + \int_{0}^{\frac{1}{2}|T|} dx\, \frac{1}{|T|} = \frac{1 + |T|^\beta}{2} \tag{18}$$

Hence,

$$W_0 = \int_{-\epsilon}^{\epsilon} dT\, \delta(T - 0^+) \int_{-\infty}^{\infty} d\beta\, \delta(\beta - |T|)\, \frac{1 + |T|^\beta}{2} \tag{19}$$

such that

$$W_0 = \int_{-\epsilon}^{\epsilon} dT\, \delta(T - 0^+)\, \frac{1 + |T|^{|T|}}{2} = \lim_{0 < T \to 0} \frac{1 + T^T}{2} = \frac{1+1}{2} = 1 \tag{20}$$

This means that when x, T, and, $\beta$, carry an index, A or B, then in the evaluation of the total density, de pdf's of those variables will result in unity in integration. This is one of the necessities of a proper total pdf. Hence, we may conclude that the densities of, x, T, and, $\beta$, with indices A or B can be introduced in a total density.

Another point of discussion is the probability description necessary for a Kolmogorovian theory. Because, T and $\theta$ follow a delta function, which is a Gaussian with infinitesimal small standard deviation, we can focus our attention to the density of x. Note that we already have shown that

$$\int_{-\frac{1}{2}|T|^\beta}^{\frac{1}{2}|T|} \frac{dx}{|T|^{\theta(x)}} = 1 \qquad (21)$$

when, $0<T\to0$, and, $\beta=|T|$. If we take $\zeta$ in the interval $[0,|T|/2]$, probabilities in that interval are between 0 and 1/2 in the limit, $0<T\to0$, and, $\theta=|T|$. In this limit, if we then also may take, $\zeta$, in $[-|T|^\theta/2,0)$. This is a uniform probability between 0 and 1/2 in the interval $[-1/2,0)$.

Note that the previous is based on

$$\int_{-\frac{1}{2}|T|^\beta}^{\xi} \frac{dx}{|T|^{\theta(x)}} = \frac{1}{|T|^{\theta(\xi)}} \left\{ \xi + \frac{1}{2}|T|^\beta \right\} \qquad (22)$$

Hence, a probability measure can be associated to the density of, x, when, $0<T\to0$. For completeness the following is given to illustrate the previous statement.

If, $\xi\geq0$, then, $\xi$, can be written as, $\xi=s|T|/2$, with $0\leq s\leq1$. The previous integral, then becomes

$$\int_{-\frac{1}{2}|T|^\beta}^{\xi} \frac{dx}{|T|^{\theta(x)}} = \frac{1}{|T|} \frac{s|T|}{2} + \frac{|T|^\beta}{2} = \frac{1}{2}(1+s) \in [0,1] \qquad (23)$$

If, $\xi<0$, then, $\xi$, can be written as, $\xi=-s|T|^\beta/2$, with $0<s\leq1$. The integral in Eq. 21 then becomes

$$\int_{-\frac{1}{2}|T|^\beta}^{\xi} \frac{dx}{|T|^{\theta(x)}} = \frac{-s|T|^\beta}{2} + \frac{|T|^\beta}{2} = \frac{|T|^\beta}{2}(1-s) \in [0,1] \qquad (24)$$

Recall that $0<T\to0$, and, $\theta=|T|$.

### 4. Definition of the behaviour of the 'bridge' between the local measurements

After definition of the probabilistic behaviour of the measuring instruments -e.g. Stern-Gerlach magnets-, the connection, or bridge, between the two measurements can be a relatively simple three-dimensional Gaussian density in $y_k$, k=1,2,3. Hence, in a simple notation,

$$N(0,1)(\vec{y}) = \frac{1}{(2\pi)^{3/2}} \exp\left[-\frac{1}{2}(y_1^2+y_2^2+y_3^2)\right] \qquad (25)$$

If we, subsequently, inspect the integrations of the product of sign functions and the three-dimensional Gaussian density, we find,

$$\int\int\int_{-\infty}^{\infty} d^3y \, N(0,1)(\vec{y}) \, sign(y_k) = 0, \qquad (k=1,2,3) \qquad \textbf{(26)}$$

because of the symmetry of the Gaussian.

Moreover, the Kronecker delta function ($\delta_{k,j}=1$, if k=j, $\delta_{k,j}=0$, if k unequal j, and k and j integer) arises in a similar manner from the product of signs and the Gaussian density. We see

$$\int\int\int_{-\infty}^{\infty} d^3y \, N(0,1)(\vec{y}) \, sign(y_k) \, sign(y_j) = \delta_{k,j} \qquad \textbf{(27)}$$

The fact that there is a bridge between the two measurements can be demonstrated from the following considerations. If the measurement functions enable the 'extraction' of the sum terms from Eq. (10), then an integration over the Gaussian combined with the sign functions in the sum terms gives

$$\int\int\int_{-\infty}^{\infty} d^3y \, N(0,1)(\vec{y}) \, \sigma_A \sigma_B =$$

$$\sum_{k=1}^{3} \sum_{j=1}^{3} a_k b_j \int\int\int_{-\infty}^{\infty} d^3y \, N(0,1) \, sign(y_k) \, sign(y_j) = \qquad \textbf{(28)}$$

$$\sum_{k=1}^{3} \sum_{j=1}^{3} a_k b_j \delta_{k,j} = \sum_{k=1}^{3} a_k b_k.$$

Hence, if the respective measurement functions can be construed such that each sum term enters a Gaussian integration, the quantum mechanical result is recovered from local hidden causality.

Of course there is the difficulty that the, to be defined, A and B measurement functions need to be either, +1, or -1, while the sum terms $\sigma_A$, and, $\sigma_B$, project in the interval [-3$^{1/2}$,3$^{1/2}$]. Hence, we may expect the integrations on the x, T, and, $\vec{\sigma}$, in relation to the A or B measurement function, to be irregular. This irregularity, on the other hand, entails that Bell's inequality cannot be derived.

## 5. The A and B measurement functions

Before introducing the measurement functions, A and B, let us recall the σ term from which the above mentioned extraction can occur. Here, we have for the A and B

$$u_{T_A}(x_A; \sigma_A) = \sigma_A \, | \, T_A \, |^{\theta(x_A)} \, \frac{d}{dx_A} \theta_{T_A}(x_A)$$

$$u_{T_B}(x_B; \sigma_B) = \sigma_B \, | \, T_B \, |^{\theta(x_B)} \, \frac{d}{dx_B} \theta_{T_B}(x_B)$$

**(29)**

Note the difference between the, $\theta(x)$, and the, $\theta_T(x)$, in the previous equation.

Writing $u_T$ more explicitly, suppressing the indices A, or, B, for the moment, we arrive at

$$u_T(x; \sigma) = \begin{cases} \sigma \, (-1)^{1-\theta(-x)} \, | \, T \, | \, \delta_T(x), & x \geq 0 \\ 3\sigma \delta_T(x), & x < 0 \end{cases}$$

**(30)**

As already has been discussed earlier (Eq. (10)), the term, $\sigma|T|\delta_T(x)$, lies in the interval [-1,1]. Ultimately, only x=0, will produce, $\lim_{0<T\to0}\{|T|\delta_T(x)\}\neq0$. Hence, the $u_T$ function can be put in a sign function which integration variable runs from -1 to 1. Let us subsequently define two uniform pdfs

$$\rho_A(\mu_A) = \begin{cases} 1/2, & \mu_A \in [-1,1] \\ 0, & \mu_A \notin [-1,1] \end{cases}$$

$$\rho_B(\mu_B) = \begin{cases} 1/2, & \mu_B \in [-1,1] \\ 0, & \mu_B \notin [-1,1] \end{cases}$$

**(31)**

Strictly speaking those two uniform densities also belong to the probabilistic behaviour of the measuring instruments, but it makes more sense to introduce them here. Given the two pdfs, we can now define the measurement functions, A and B.

$$A = sign\{u_{T_A}(x_A; \sigma_A) - \mu_A\}$$

$$B = -sign\{u_{T_B}(x_B; \sigma_B) - \mu_B\}$$

**(32)**

The extraction of, $\sigma_A$, can now be demonstrated. In the first place, we may write in the evaluation of the statistical expectation of the function A,

$$\frac{1}{2}\int_{-1}^{1} d\mu_A \, sign\{u_{T_A}(x_A; \sigma_A) - \mu_A\} = u_{T_A}(x_A; \sigma_A)$$

**(33)**

Later in the paper a more precise way of writing down the 'extraction plus bridge' process of integration will be given. If however, we accept for the moment that $u_T$ can be obtained this way, then we may be interested in the role of the, $x_A$, $T_A$, and $\theta_A$ hidden variables. Hence, we are interested in

$$W_1 =$$

$$\int_{-\infty}^{\infty} dT_A \int_{-\infty}^{\infty} d\beta_A \int_{-\infty}^{\infty} dx_A \, \rho_{T_A}(T_A) \, \rho_{B_A}(\beta_A) \, \rho_{X_A}(x_A; T_A, \beta_A) \, u_{T_A}(x_A; \sigma_A) \; . \tag{34}$$

For the auxiliary variable, $W_1$, we firstly integrate over $x_A$. Hence,

$$\int_{-\infty}^{\infty} dx_A \, \rho_{X_A}(x_A; T_A, \beta_A) \, u_{T_A}(x_A; \sigma_A) =$$

$$\int_{-\frac{1}{2}|T_A|^{\beta_A}}^{\frac{1}{2}|T_A|} dx_A \, \frac{u_{T_A}(x_A; \sigma_A)}{|T_A|^{\theta(x_A)}} \tag{35}$$

Furthermore, we see from the form, $u_T = \sigma |T|^{\theta(x)} \delta_T(x)$, that

$$\int_{-\frac{1}{2}|T_A|^{\beta_A}}^{\frac{1}{2}|T_A|} dx_A \, \frac{u_{T_A}(x_A; \sigma_A)}{|T_A|^{\theta(x_A)}} = \sigma_A \int_{-\frac{1}{2}|T_A|^{\beta_A}}^{\frac{1}{2}|T_A|} dx_A \, \frac{d}{dx_A} \, \theta_{T_A}(x_A) = \tag{36}$$

$$\sigma_A \{ \theta_{T_A}(\frac{1}{2}|T_A|) - \theta_{T_A}(-\frac{1}{2}|T_A|^{\beta_A}) \}$$

Subsequently, this implies for $W_1$,

$$W_1 =$$

$$\sigma_A \int_{-\epsilon}^{\epsilon} dT_A \int_{-\infty}^{\infty} d\beta_A \{ \delta(T_A - 0^+) \, \delta(\beta_A - |T_A|) \tag{37}$$

$$\times \{ \theta_{T_A}(\frac{1}{2}|T_A|) - \theta_{T_A}(-\frac{1}{2}|T_A|^{\beta_A}) \} \} \; .$$

Further evaluation gives

$$W_1 = \sigma_A \int_{-\epsilon}^{\epsilon} dT_A \, \delta(T_A - 0^+) \, \{ \theta_{T_A}(\frac{1}{2}|T_A|) - \theta_{T_A}(-\frac{1}{2}|T_A|^{|T_A|}) \} = \tag{38}$$

$$\sigma_A \lim_{0 < T_A \to 0} \{ \theta_{T_A}(\frac{1}{2}|T_A|) - \theta_{T_A}(-\frac{1}{2} \, T_A^{T_A}) \} = \sigma_A \{ \theta(0) - \theta(-\frac{1}{2}) \}$$

Hence, $W_1 = \sigma_A$. This result shows that $\sigma_A$ can be extracted from the A measurement function. Similar result can be obtained from the B measurement function. Moreover, we see that the integration is irregular because the $u_T$ lies in [-1,1], while, its result, $\sigma_A$, can be larger than 1 and smaller than -1. Hence, irregular integration is the reason that Schwarz's inequality in certain phases of the evaluation of the expectation of A, of B, and/or of AB, fails.

Let us also show in more detail why the Bell inequality cannot be obtained from the model.

Firstly, we note that generally for a model of this type, we have for the covariance of A and B

$$\langle AB \rangle = \int d\lambda \, \rho_1(\lambda) \int d\mu \, \rho_2(\mu) \int d\nu \, \rho_3(\nu) \, A(\lambda, \mu, \vec{a}) \, B(\lambda, \nu, \vec{b}) \qquad \textbf{(39)}$$

where, $\lambda$, $\mu$, $\nu$, denote groups of variables. Secondly, we note that Bell's inequality depends on Schwarz's inequality. Hence, when A=+1, or, -1, and similar for B, we write

$$|\langle AB \rangle| \leq \int d\lambda \, \rho_1(\lambda) \int d\mu \, \rho_n(\mu) \int d\nu \, \rho_3(\nu) \, |A(\lambda, \mu, \vec{a})| \, |B(\lambda, \nu, \vec{b})| \qquad \textbf{(40)}$$

This then, in a regular case, leads to, |AB|≤1, hence, the absolute covariance between A and B is less than or equal to unity.

If, however, this inequality is valid, we also must have

$$|\langle AB \rangle|$$
$$\leq \int d\lambda \, \rho_1(\lambda) \, \left| \int d\mu \, \rho_2(\mu) \, A(\lambda, \mu, \vec{a}) \right| \, \left| \int d\nu \, \rho_3(\nu) \, B(\lambda, \nu, \vec{b}) \right| \qquad \textbf{(41)}$$

In our case we can identify

$$|W_1| =$$

$$\left| \int\limits_{-e}^{e} dT \, \delta(T - 0^+) \int\limits_{-\infty}^{\infty} d\beta \, \delta(\beta - |T|) \int\limits_{-\frac{1}{2}|T|^\beta}^{\frac{1}{2}|T|} \frac{dx}{2 \, |T|^{\theta(x)}} \int\limits_{-1}^{1} d\mu \, sign(u_T - \mu) \right| . \qquad \textbf{(42)}$$

Here, $W_1 = W_1(A)$, is an expression in terms of the model, of the first term between absolute lines on the right hand of Eq. (40). A similar remark applies to $W_1(B)$, which is the second term between absolute marks. In a regular case, $|W_1| \leq 1$, however, if we evaluate like it is done in the previous sections, it follows

$$| W_1 | =$$

$$\left| \int_{-\epsilon}^{\epsilon} dT \delta (T-0^+) \int_{-\infty}^{\infty} d\beta \, \delta (\beta - | T |) \int_{-\frac{1}{2} | T |^{\beta}}^{\frac{1}{2} | T |} \frac{dx}{2 | T |^{\theta(x)}} \int_{-1}^{1} d\mu \; sign(u_T - \mu) \right| =$$

$$\left| \int_{-\epsilon}^{\epsilon} dT \delta (T-0^+) \int_{-\infty}^{\infty} d\beta \, \delta (\beta - | T |) \int_{-\frac{1}{2} | T |^{\beta}}^{\frac{1}{2} | T |} \frac{dx}{| T |^{\theta(x)}} u_T(x;\sigma) \right| = \qquad \textbf{(43}$$

$$\left| \sigma \int_{-\epsilon}^{\epsilon} dT \delta (T-0^+) \int_{-\infty}^{\infty} d\beta \, \delta (\beta - | T |) \int_{-\frac{1}{2} | T |^{\beta}}^{\frac{1}{2} | T |} dx \frac{d}{dx} \theta_T(x;\sigma) \right| =$$

$$= | \sigma | .$$

The previous expression and,$|W_1| \leq 1$, are contradictory because, for, $a_1=a_2=a_3=(1/3)^{1/2}$, $(a_1^2+a_2^2+a_3^2=1)$, we can have, $|\sigma|>1$. Hence, Schwarz's inequality cannot be applied. Hence no Bell inequality exists for this type of $\langle AB \rangle$.

Note that, $\sigma|T|\delta_T(x)$, or $u_T$, projects in the interval [-1,1]. Note also, as a rule of the thumb, that $|T|$ in its role as standard deviation gives, Prob$[-3|T| \leq x \leq 3|T|] \approx 0.99$. Hence, 99% of the surface below the Gaussian lies between -3$|T|$ and 3$|T|$. Note in addition also that in Generalized function theory, x and T are independent variables[6], hence, no problem arises when interchanging limit with integration.

## 6. The complete model and the evaluation of the expectation

The complete pdf in the model is

$$\rho_{A-wing} = \rho_A(\mu_A) \rho_{T_A}(T_A) \rho_{\beta_A}(\beta_A) \rho_{X_A}(x_A; T_A, \beta_A)$$

$$\rho_{B-wing} = \rho_B(\mu_B) \rho_{T_B}(T_B) \rho_{\beta_B}(\beta_B) \rho_{X_B}(x_B; T_B, \beta_B) \qquad \textbf{(44)}$$

$$\rho_{tot} = \rho_{A-wing} \rho_{B-wing} N(0,1)(\vec{y})$$

In the evaluation of the statistical expectations of the model, we are in need of a notation to denote the ongoing integration process. Let us continue to use the previously loosely defined bracket notation, which is resembles the one from statistical physics. An already partly worked-out example on the expectation of A can now be used to show the workings of the bracket notation we have.

$$\langle A \rangle = \langle u_{T_A}(x_A; \sigma_A) \rangle \qquad \textbf{(45)}$$

Because, A=sign($u_{TA}-\mu_A$), which after integration over, $\mu_A$, gives the above result.

In the employed notation the angular brackets indicate the ongoing integration. In the previous, the pdf weighted integration over, $\mu_A$, has been performed, such as in Eq. (28). Subsequent pdf weighted integration over the, $x_A$, $T_A$, and $\boldsymbol{\theta}_A$ hidden variables, such as in Eqs.(34)-(38), then gives

$$\langle A \rangle = \langle \sigma_A \rangle =$$
$$\int\limits_{-\infty}^{\infty}\int\int d^3 y\, N(0,1)(\vec{y})\, \sigma_A = 0 \tag{46}$$

The vanishing of $\langle A \rangle$ follows from Eq. (26). Similar result follows for $\langle B \rangle = 0$.

Moreover, with the similar notation, it easily follows that, $\langle 1 \rangle = \langle A^2 \rangle = \langle B^2 \rangle = 1$. Conclusively, the result for the covariance of A and B can be written as

$$P(\vec{a}, \vec{b}) = \langle AB \rangle =$$
$$-\langle u_{T_A}(x_A; \sigma_A)\, u_{T_B}(x_B; \sigma_B)\rangle = -\langle \sigma_A \sigma_B \rangle \tag{47}$$

This result arises in a similar manner as, $\sigma_A$, from the expectation of A, because the pdf weighted integration over, $x_A$, $T_A$, and $\boldsymbol{\theta}_A$, is independent from, weighted integration over, $x_B$, $T_B$, and $\boldsymbol{\theta}_B$, and, AB=-sign($u_{TA}$-$\mu_A$)sign($u_{TB}$-$\mu_B$). From the previous equation, for completeness, it then follows that

$$P(\vec{a}, \vec{b}) =$$
$$\frac{-1}{(2\pi)^{3/2}} \sum_{k=1}^{3} \sum_{j=1}^{3} a_k b_j \int\limits_{-\infty}^{\infty}\int\int d^3 y\, N(0,1)(\vec{y})\, sign(y_k)\, sign(y_j) \tag{48}$$

This results into the following expression.

$$P(\vec{a}, \vec{b}) = -\sum_{k=1}^{3} a_k b_k. \tag{49}$$

### 7. Conclusion & discussion

The main conclusion is that the quantum result for the covariance of A and B can be obtained from local hidden causality.

Finally, the derivation of separate probabilities for, A=+1, and for, A=-1 is discussed. Let us inspect the P(A=+1). The question is, is this probability equal to the probability P(A=-1), that is, is P(A=+1)=P(A=-1)=1/2 ?

In the first place we note that A=+1, only when, $u_T > \mu_A$. The associated uniform probability then is equal to

$$P_{A=+1}(u_{T_A} > \mu_A) = \left\{ \begin{array}{ll} \frac{1}{2}\left(1 + \frac{\sigma_A}{\sqrt{2\pi}}\right), & x=0 \\ \\ \frac{1}{2}, & x \neq 0 \end{array} \right.$$

The other variables, save the, $y_k$, (k=1,2,3) indirectly via, $\sigma_A$, do not influence the sign of A. Hence, we may conclude that the other variables run through the complete intervals as defined previously. We then see

$$P(A=+1) = \int_{A=+1} \rho_{tot} = 1/2.$$

P(A=−1)=1/2 follows similarly.

It can therefore rightfully be claimed that the Kolmogorovian proposed model refutes the theorem that Bell's inequalities prevent local hidden variables to recover quantum covariance. Furthermore, it refutes the notion that no fruitful, classical type, local hidden variable theory is possible. Perhaps that there are no local hidden causalities, but that is unrelated to Bell's theorem. Moreover, because a working model has been created, arguments against LHV theories without Bell inequalities also fail. In conclusion, it can rightfully be claimed that quantum mechanics is not the exclusive explanation for the experimentally obtained quantum covariance.